\documentclass[aps,pra,amsmath,showpacs,reprint]{revtex4-1}

\usepackage{hyperref}
\usepackage[T1]{fontenc}
\usepackage[utf8]{inputenc}
\usepackage{rotating}
\usepackage{braket}
\usepackage[capitalise]{cleveref}
\usepackage{amsmath,amssymb,amsfonts, bm,amsthm, dsfont} %Mathematical Symbol Packages
\usepackage{graphicx} %Figures Packages
\usepackage{color}
\usepackage{dsfont}
\usepackage[english]{babel}

\newcommand{\half}{\frac{1}{2}}
\newcommand{\Hilb}{\mathcal{H}}
\newcommand{\Liou}{L}

\newcommand{\K}{\mathcal{K}}
\newcommand{\Qproj}{\mathcal{Q}}
\newcommand{\Pproj}{\mathcal{P}}
\newcommand{\Lprop}{\mathcal{L}}
\newcommand{\Gprop}{\mathcal{G}}
\DeclareMathOperator{\Tr}{Tr}
\DeclareMathOperator{\TrB}{Tr_B}

\DeclareMathOperator{\dive}{div}

\begin{document}
\date{\today}
\flushbottom

\title{
State space distribution and dynamical flow for closed and open quantum systems
} 
\author{Amro Dodin}
\affiliation{Department of Chemistry, Massachusetts Institute of Technology, Cambridge, Massachusetts, 02139, USA}
\author{Adam P. Willard}
\affiliation{Department of Chemistry, Massachusetts Institute of Technology, Cambridge, Massachusetts, 02139, USA}

\begin{abstract}
We present a general formalism for studying the effects of dynamical heterogeneity in open quantum systems.
We develop this formalism in the state space of density operators, on which ensembles of quantum states can be conveniently represented by probability distributions.
We describe how this representation reduces ambiguity in the definition of quantum ensembles by providing the ability to explicitly separate classical and quantum sources of probabilistic uncertainty.
We then derive explicit equations of motion for state space distributions of both open and closed quantum systems and demonstrate that resulting dynamics take a fluid mechanical form analogous to a classical probability fluid on Hamiltonian phase space, thus enabling a straightforward quantum generalization of Liouville's theorem.
We illustrate the utility of our formalism by analyzing the dynamics of an open two-level system using the state-space formalism that are shown to be consistent with the derived analytical results.
\end{abstract}

\maketitle

% =================================================================================================================
% Introduction
% =================================================================================================================
One of the foundational concepts in physics is that the observable properties of a macroscopic systems can be represented as an average over an ensemble of identical but statistically independent microscopic subsystems.
The field of statistical mechanics provides the theoretical formalism for characterizing these subsystem statistics and relating them to macroscopic observables.
In this formalism, ensemble statistics and dynamics are conveniently expressed in terms of time evolving probability distributions over the subsystem state space.
In classical mechanics, these probability distributions are formulated in Hamiltonian phase space and evolve according to Liouville's Equation.
Generalizing this formulation to quantum mechanics has been a longstanding problem due to the wave nature of quantum states, which does not allow for a well defined probability distribution in phase space.
In this letter, we show that this problem can be solved by expressing quantum ensembles as probability distributions over he state space of density operators instead.

The state of a quantum ensemble is typically described by a single density operator, $\hat{\rho}$.
This representation efficiently encodes the statistics of ensemble observables and is central to quantum theories for 
dynamics \cite{blum_density_2012,breuer_theory_2007,feynman_theory_2000,makri_numerical_1995}, 
optics \cite{mandel_optical_1995,scully_quantum_1997}, 
thermodynamics \cite{alicki_quantum_2014,alicki_quantum_1979,parrondo_thermodynamics_2015}, 
information \cite{nielsen_quantum_2010,kaye_introduction_2007} 
and control \cite{shapiro_quantum_2012,dong_quantum_2010,wiseman_quantum_2009}.
However, a single density operator is an incomplete description of an ensemble because it contains no information about the state of individual systems.
Rather, $\hat{\rho}$ describes only the uncertainty of the ensemble observables, combining classical contributions, due to an unknown initial state of the subsystems, and quantum contributions, due to the random outcome of measurements on superposition states.
As a result, different ensembles can be represented by the same $\hat{\rho}$, complicating their microscopic interpretation.
The importance of this physical insight has been exemplified in the study of light-induced biomolecular dynamics \cite{brumer_shedding_2018,pachon_open_2017,dodin_coherent_2016,dorfman_photosynthetic_2013,kassal_does_2013,brumer_molecular_2012}, 
environment-conditioned qubit dynamics \cite{ficheux_dynamics_2018},
and emerging ultrafast single-molecule spectroscopies \cite{weigel_shaped_2015,brinks_ultrafast_2014,hildner_femtosecond_2011,brinks_visualizing_2010,hernando_effect_2006,van_dijk_single-molecule_2005}.

Here, we present a theoretical formalism for treating quantum ensembles analogously to classical ensembles.
This approach resolves the challenges that arise due to the wave properties of quantum states on phase space by working in a natural quantum state space. 
By considering stochastic processes on Liouville space, as first proposed by Davies \cite{davies_quantum_1969}, we define a quantum state space probability density.
We then derive equations of motion for these distributions, which remarkably take the same form as the classical Liouville's Equation.
This equivalence systematically generalizes the methods and intuition of classical statistical mechanics to quantum systems, which we illustrate by proving a novel quantum Liouville Theorem.

% =================================================================================================================
% Formulation
% =================================================================================================================
\noindent
\textit{\textbf{Classical ensembles on phase space:}}

\begin{figure}[htbp]
  \includegraphics[width=\columnwidth]{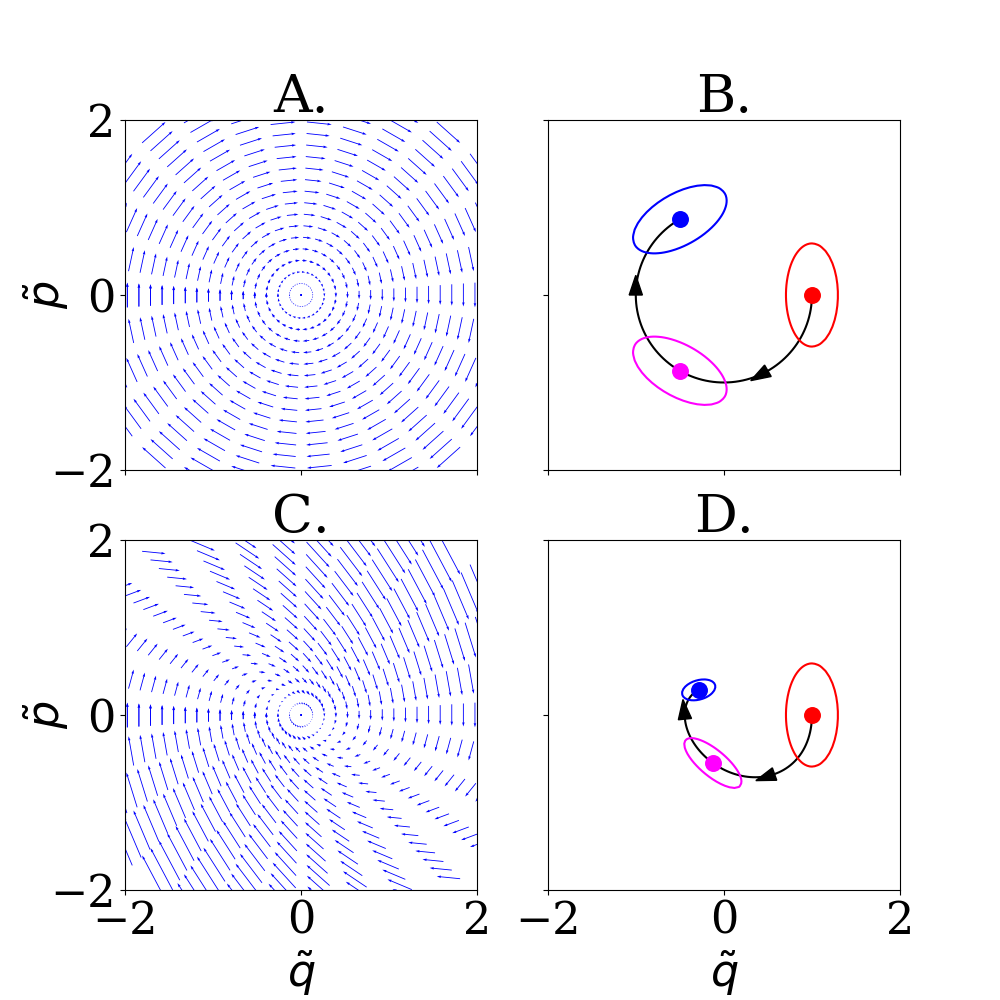}
  \caption{Classical flow fields for simple (A) and linearly damped (C) harmonic oscillators.  Sample trajectories (black traces) and distribution isosurfaces (colored contours) for these models are shown in (B) and (D). Red, pink and blue contours and points are evaluated at $t=0, 2\pi/3\omega$ and $4\pi/3\omega$ respectively, where $\omega$ is the undamped oscillator frequency. Damping coefficient is $\gamma = \omega/2$. Results shown in symmetrized coordinates $\tilde{p}:= p/\sqrt{m\omega}$, $\tilde{q} := q\sqrt{m\omega}$ where $m$ is the particle mass.}
  \label{fig:c}
\end{figure}

We begin by briefly outlining the classical theory of phase space ensembles.
The state of a classical system is described by enumerating the positions and momenta of all particles.
For a system with $N$ position coordinates $\lbrace q_i\rbrace^N_{i=1}$  and momenta $\lbrace p_i\rbrace^N_{i=1}$, this defines a $2N$-dimensional vector $\bm{x} = [\bm{q}, \bm{p}]$ in phase space, $\mathbb{R}^{2N}$.
The evolution of a closed system is governed by Hamilton's equations of motion and defines a vector field $\dot{\bm{x}} = [\partial H/\partial\bm{q}, -\partial H/\partial\bm{p}]$ called the dynamical flow field.
This can be modified to treat dissipative systems by adding damping terms.
For example, linear dissipators, given by positive matrices $\lbrace\hat{\Gamma}_\alpha \rbrace$ yield a dissipative flow field, $\dot{\bm{x}} = \dot{\bm{x}}_C - \sum_\alpha\hat\Gamma_\alpha \bm{x}$ where  $\dot{\bm{x}}_C$ is the closed system flow field.
A system initially in $\bm{x}_0$ propagates along the flow field to $\bm{x}(t|\bm{x}_0)$ at time $t$, tracing out a trajectory $\lbrace \bm{x}(t| \bm{x}_0)|t\geq0 \rbrace$.
Example flow fields are shown for closed and damped harmonic motion in Fig. \ref{fig:c}.A and C.

An ensemble is comprised of a collection of systems, each found in a different state.
The state of an ensemble is then given by a probability distribution $P(\bm{x}; t)$ where $\bm{x}$ is a coordinate identifying a point in phase space.
Liouville's equation describes the dynamics of this distribution in terms of the flow field, giving
\begin{equation}
  \label{eq:c_liou}
  \frac{\partial P}{\partial t}(\bm{x}; t) = -\bm{\nabla} P(\bm{x}; t) \cdot \dot{\bm{x}} + \kappa(\bm{x}) P(\bm{x}; t),
\end{equation}
where $\kappa(\bm{x}) := -\dive(\dot{\bm{x}})$ is the compressibility of the flow field and $\bm{a}\cdot\bm{b} := \sum_{i=1}^{2N} a_i b_i$ is the inner product on $\mathbb{R}^{2N}$.
If the single system dynamics, $\bm{x}(t|\bm{x}_o)$, are known, Eq. (\ref{eq:c_liou}) can be analytically solved to give $P(\bm{x}; t) = \int d\bm{x'} P(\bm{x'}; 0) \delta(\bm{x}(t| \bm{x'}) - \bm{x})$, where the convection Green's Function $\delta(\bm{x}(t| \bm{x'}) - \bm{x})$ selects initial conditions $\bm{x'}$ that evolve to  $\bm{x}$ at time $t$.
Distribution  and trajectory dynamics are illustrated for two harmonic oscillator in Fig. \ref{fig:c}.B and D.

The isosurfaces in Fig. \ref{fig:c} reveal an important property of classical systems.
Closed system distributions remain the same size at all times, while damped system distributions compress towards a low energy steady state.
Liouville's theorem relates this notion of phase space volume to the reversibility and determinism of the underlying dynamics.
In a closed system, the phase space volume, or equivalently the local density evaluated along an evolving trajectory, $P(\bm{x}(t|\bm{x}_0); t)$, is constant and the flow field is incompressible everywhere ($\kappa(\bm{x})=0$), reflecting the fact that reversible, deterministic trajectories cannot cross.
This property does not hold for open systems where linear dissipative flow has positive compressibility.
The distribution collapses onto a steady state point $\bm{x}_{SS}$ where the dynamics are irreversible since the initial state of a trajectory that has relaxed to $\bm{x}_{SS}$ cannot be determined.

\noindent
\textit{\textbf{Quantum ensembles in Liouville space:}}

This classical statistical structure is effective because it is constructed on a space where each point uniquely identifies a state.
Motivated by this observation, we formulate quantum ensembles on a quantum state space rather than on classical phase space.
The state of an open quantum system is defined by a density operator $\hat{\sigma}$ in Liouville space, $\Liou(\Hilb)$, a complex vector space of linear operators acting on the system Hilbert space, $\Hilb$.
Given a basis $\lbrace \ket{i} \rbrace_{i=1}^N$ for $\Hilb$, any operator can be written in vector form $\hat{\sigma} \to \bm{\sigma} := [\sigma_{11}, \sigma_{12}, ..., \sigma_{NN}]$ by listing its matrix elements $\sigma_{ij}:= \bra{i}\hat\sigma \ket{j}$.
Moreover, it is equipped with a \textit{trace inner product}, $\bm{\alpha}\cdot\bm{\beta} := \Tr\lbrace\hat{\alpha}^\dagger\hat{\beta}\rbrace = \sum_{ij}\alpha_{ij}^*\beta_{ij}$.

Similarly to classical ensembles, quantum ensembles are comprised of a collection of systems, each in a different state.
The state of the ensemble can then be defined by a probability distribution $P(\bm{\sigma}; t)$ on Liouville space where $\bm{\sigma}$ is a coordinate that identifies a point in Liouville space.
This formulation provides a more detailed account of the state of a quantum ensemble than the ensemble density matrix $\hat\rho$.
While $\hat\rho$ describes only the statistics of the ensemble observables, $P(\bm{\sigma}; t)$ encodes the state  of subsystems in the ensemble.
In fact, $\hat{\rho}$ is the average of this distribution, $\bm{\rho}(t) := \int d\bm{\sigma} P(\bm{\sigma}; t) = \langle \bm{\sigma} \rangle (t)$, that integrates over the classical uncertainty.
As a result, $P(\bm{\sigma}; t)$ resolves the ambiguity between quantum and classical uncertainty by treating each quantum state as a well-defined point.
The classical uncertainty in the preparation of system states is described by the probability distribution, $P(\bm{\sigma}; t)$.
The quantum uncertainty in observable outcomes of each subsystem is separately contained in the quantum description of it's state, $\bm{\sigma}$.

% =================================================================================================================
% Dynamics
% =================================================================================================================
\noindent
\textit{\textbf{Quantum dynamical flow on Liouville space:}}

We now consider the dynamics of an ensemble of $N$ dimensional quantum systems.
First, we define a quantum dynamical flow field on Liouville space.
This describes the evolution of a quantum system at point $\bm{\sigma}$.
For a closed system with Hamiltonian $\bm{H}$, this is given by the Liouville von-Neumann equation:
\begin{equation}
  \label{eq:LvN}
  \dot{\bm{\sigma}} = \Lprop_0\bm\sigma,
\end{equation}
where $\Lprop_0=-\frac{i}{\hbar}[\bm{H}, \cdot]$ is the Liouville superoperator.

This can be extended to open systems by adding dissipator terms.
For simplicity, we restrict our attention in the main text to Markovian dynamics, providing the general non-Markovian Nakajima-Zwanzig theory \cite{nakajima_quantum_1958,zwanzig_ensemble_1960,breuer_theory_2007} in the SI.
The general form of Markovian dynamics is given by the the Gorini-Kossakowski-Sudarshan-Lindblad (GKSL) equation \cite{gorini_completely_1976,lindblad_generators_1976,alicki_quantum_2007,breuer_theory_2007,blum_density_2012},
\begin{equation}
  \label{eq:GKSL}
  \dot{\bm{\sigma}} = \Lprop \bm{\sigma} := \Lprop_0(t) \bm{\sigma} + \sum_{\alpha=1}^{N^2-1} \gamma_{\alpha} \mathcal{D}[\bm{L}_\alpha]\bm{\sigma},
\end{equation}
where the Lindblad operators $\lbrace \bm{L}_\alpha \rbrace_{\alpha=1}^{N^2}$ form an orthonormal basis of $\Liou(\Hilb)$, $\mathcal{D}[\bm{L}_\alpha] \bm{\sigma} = \bm{L}_\alpha \bm{\sigma} \bm{L}_\alpha^\dagger - \half \lbrace \bm{L}_\alpha^\dagger \bm{L}_\alpha, \bm{\sigma} \rbrace$ is the associated dissipators with rates $\gamma_\alpha \geq 0$ and $\lbrace\bm{A}, \bm{B}\rbrace = \bm{A}\bm{B} + \bm{B}\bm{A} $ is the anticommutator.
By convention $\bm{L}_{N^2} = \bm{\mathbb{I}}$ is the identity operator and $\lbrace \bm{L}_\alpha \rbrace_{\alpha=1}^{N^2 -1}$ have vanishing trace.
This expression takes a form similar to classical linear dissipation described in the previous section.

A system initialized in state $\bm{\sigma}_0$, evolves for a time $t$ by moving along the flow field.
This yields a state $\bm{\sigma}(t| \bm{\sigma_0})$ and traces out a trajectory $\lbrace \bm{\sigma}(t|\bm{\sigma}_0)|t\geq 0\rbrace$ through Liouville space that is continuous and defined at all times.
Since trajectories are defined at all times, dynamics never creates or destroys trajectories.
This property is known as trajectory conservation.

Using these properties, we derive a continuity equation for trajectories in Liouville space.
Since trajectories are continuous and conserved, all trajectories entering or leaving a region, $\Omega$, must pass through its boundary.
Therefore, the change in the probability $P_\Omega$ of finding a trajectory in $\Omega$ is related to the probability flux $\bm{j}(\bm{\sigma}) := P(\bm{\sigma}) \dot{\bm{\sigma}}$ passing through its boundary.
This can be written in differential form as,
\begin{equation}
  \label{eq:continuity}
  \frac{\partial P}{\partial t} (\bm{\sigma}; t)= -\dive(\bm{j}) = -\bm{\nabla}P(\bm{\sigma}; t) \cdot \dot{\bm{\sigma}} + P(\bm{\sigma}; t)\kappa(\bm{\sigma}),
\end{equation}
where $\kappa(\bm{\sigma}):= -\dive(\dot{\bm{\sigma}})$ is the flow field compressibility, the gradient of a function $f(\bm{\sigma})$ is $\bm{\nabla}f(\bm{\sigma}) := [\partial f(\bm{\sigma})/\partial \sigma_{11}, ..., \partial f(\bm{\sigma})/ \partial \sigma_{NN}]$, and the divergence of a vector field $\bm{g}(\bm{\sigma})$ is $\dive(\bm{g}(\bm{\sigma})) := \sum_{ij}\partial g_{ij}(\bm{\sigma})/\partial\sigma^*_{ij}$.
This derivation mirrors the classical phase space continuity equation and is detailed in the SI.
\footnote{We note that, while these complex differential operations can in principle yield complex valued results, they are guaranteed to be real when restricted to the subspace of self-adjoint linear operators that contains all system density matrices.}
Equation (\ref{eq:continuity}) can be applied to closed and open dynamics by substituting the appropriate flow field (Eq. (\ref{eq:LvN}) or (\ref{eq:GKSL})) to give,
\begin{subequations}
  \label{eqs:distdyn}
  \begin{equation}
    \label{eq:LvnDist}
    \frac{\partial P}{\partial t} = \bm{\nabla}P(\bm{\sigma}; t) \cdot \Lprop_0 \bm{\sigma},
  \end{equation}
  \begin{equation}
    \label{eq:GKSLDsit}
     \frac{\partial P}{\partial t} = \bm{\nabla}P(\bm{\sigma}; t) \cdot \Lprop \bm{\sigma} + P(\bm{\sigma}; t)N\sum_{\alpha=1}^{N^2-1}\gamma_\alpha.
  \end{equation}
\end{subequations}

Remarkably, Eq. (\ref{eq:continuity}) describing quantum ensemble dynamics takes an identical form to  Eq. (\ref{eq:c_liou}) for classical ensembles, thus revealing that uncertainty in the initial preparation of systems in an ensemble propagates similarly for quantum and classical systems.
Consequently, the problematic behavior of quantum mechanics on phase space arises due to the incompatibility of quantum uncertainty, that prohibits the simultaneous knowledge of positions and momenta, with a position momentum state space.
This difficulty can be avoided by treating quantum dynamics on its natural state space where the statistical structure of a dynamical system is conserved.

The similarity between Eq. (\ref{eq:c_liou}) and (\ref{eq:continuity}) makes it straightforward to propagate the dynamics of quantum ensembles.
Methods used to solve Eq. (\ref{eq:c_liou}) can be directly applied to Eq. (\ref{eq:continuity}).
If the dynamics of the microscopic system, $\bm{\sigma}(t| \bm{\sigma_0})$ are known, the distribution dynamics can be obtained using the Green's function method, yielding
\begin{equation}
  \label{eq:gfsol}
  P(\bm{\sigma}; t) = \int d\bm{\sigma'} P(\bm{\sigma}'; 0) \delta(\bm{\sigma}(t|\bm{\sigma}') - \bm{\sigma}),
\end{equation}
where the Green's Function $\delta(\bm{\sigma}(t|\bm{\sigma}') - \bm{\sigma})$ selects initial conditions $\bm{\sigma}'$ that evolve to state $\bm{\sigma}$ at time $t$.

\noindent
\textit{\textbf{Quantum Liouville Theorem}}

The identical structure of quantum and classical ensembles can also be exploited to systematically generalize key classical results built on Liouville's equation.
To provide a road map for extending classical results we show that the classical Liouville's theorem can be trivially extended to quantum ensembles.
Notably, previous phase space studies \cite{kakofengitis_wigners_2017,skodje_flux_1989, price_quantum_1995} have concluded that quantum dynamical flow in phase space cannot be written in an incompressible form.
This shows that quantum dynamics on phase space differs fundamentally from classical mechanics.
In contrast, the applicability of Liouville's theorem shows that classical notions of reversibility and determinism can still be applied to quantum ensembles on $\Liou(\Hilb)$.

Following the classical derivation, we consider the probability density evaluated along a Liouville space trajectory $P(\bm{\sigma}(t|\bm{\sigma_0}); t)$.
The rate of change of this quantity is 
\begin{equation}
  \label{eq:localP}
  \frac{d P}{dt}(\bm{\sigma}(t| \bm{\sigma}_0); t) =  \kappa(\bm{\sigma}(t| \bm{\sigma}_0))P(\bm{\sigma}(t|\bm{\sigma_0}); t), 
\end{equation}
after an application of the chain rule and a substitution of Eq. (\ref{eq:continuity}).
By substituting the Liouville-von Neumann flow field for a closed system (Eq. (\ref{eq:LvN})), we find that $\dive(\dot{\bm{\sigma}}) = 0$ for all $\bm{\sigma}$ since $\dive(\bm{H}\bm{\sigma}) = \dive(\bm{\sigma}\bm{H})$.
This indicates that the classical picture of deterministic, reversible dynamics in terms of non-intersecting state space trajectories applies directly to quantum trajectories.
Moreover, the derivation of this result directly mimics the classical proof, highlighting the ease of applying classical derivations in this formalism.
The same process can be repeated for GKSL dynamics.
Substituting Eq. (\ref{eq:GKSL}) into Eq. (\ref{eq:localP}) gives a uniform compressibility $\kappa(\bm{\sigma}) = N\sum_{\alpha=1}^{N^2-1}\gamma_\alpha \geq 0$, taking a similar form to classical linear dissipation.
A glossary summarizing the quantum-classical analogies in our formalism is provided in the SI.

% =================================================================================================================
% Spin Boson Example
% =================================================================================================================
\noindent
\textit{\textbf{Spin-Boson Distribution Dynamics}}

To demonstrate the application of the formulation presented above we consider the dynamics of a two-level spin system. 
We describe density matrices using the Pauli basis $\lbrace \bm{S}_i \rbrace_{i=0, x, y, z}$ in $\Liou(\Hilb)$,
  where $\bm{S}_0 := \bm{\mathbb{I}}/\sqrt{2}$ is the normalized identity operator, and $\bm{S}_x := (\ket{1}\bra{2} + \ket{2}\bra{1})/\sqrt{2}$, $\bm{S}_y := -i(\ket{1}\bra{2} - \ket{2}\bra{1})/\sqrt{2}$, and $\bm{S}_z := (\ket{2}\bra{2} -\ket{1}\bra{1})/\sqrt{2}$.
 Any density matrix $\bm{\sigma} = (\bm{S}_0 + x \bm{S}_x + y \bm{S}_y + z \bm{S}_z)/\sqrt{2}$ can be written as a real valued 3D vector $\bm{\sigma} \to [x, y, z]$ with $\sqrt{x^2 + y^2 +z^2}\leq 1$.
 
\begin{figure}[h]
  \centering
  \includegraphics[width=\columnwidth]{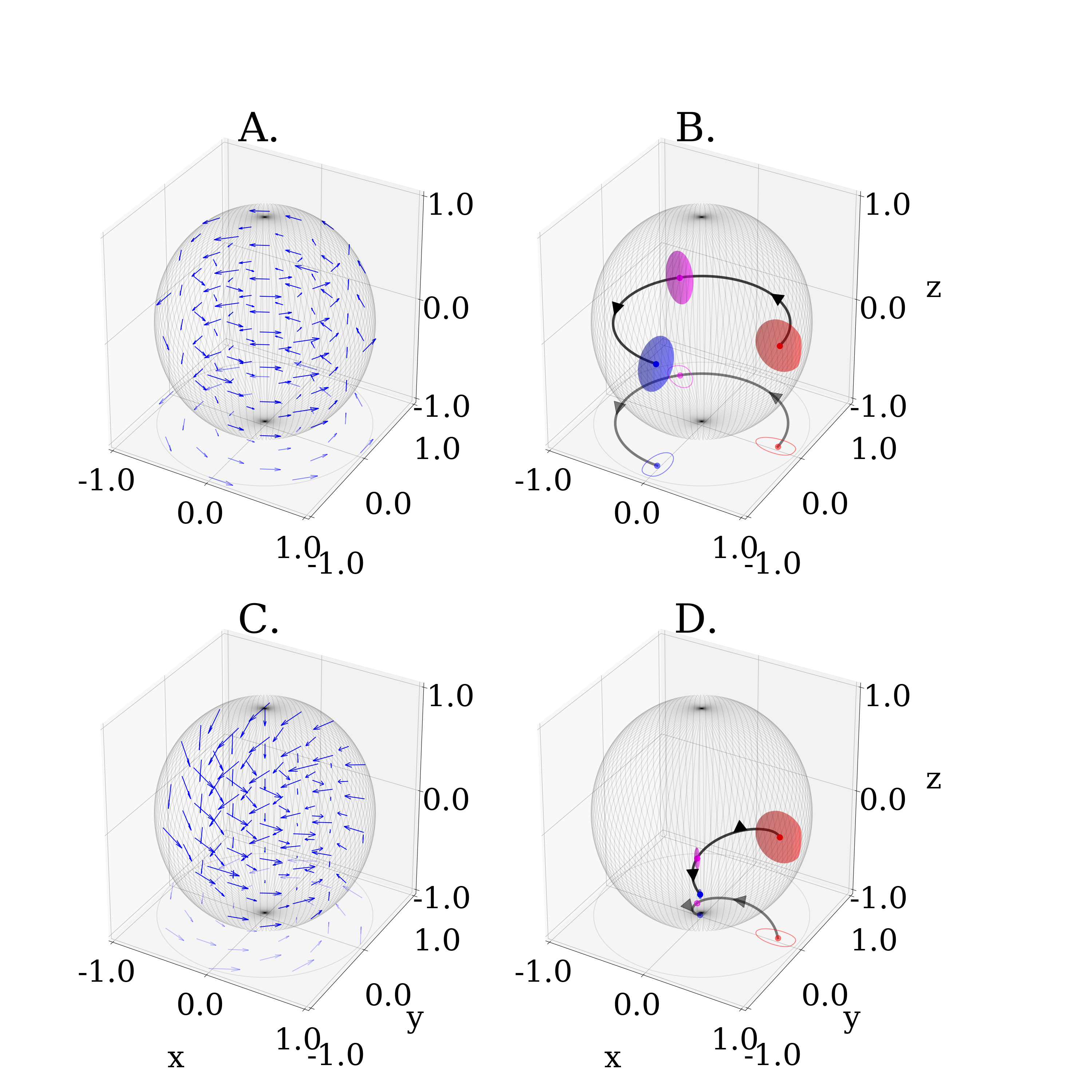}
  \caption{Quantum flow fields for isolated spin (A) and spin-boson (C) models.  Sample trajectories (black traces) and isosurfaces (colored contours)  for these models are shown in (B) and (D). Red, pink and blue contours and points are evaluated at $t=0, 2\pi/3\omega$ and $4\pi/3\omega$ where $\hbar\omega$ is the spin energy splitting. Lindblad rates are $\gamma_\phi =\omega/3=\Gamma$.}
  \label{fig:q}
\end{figure}

 Consider the Markovian dynamics of a spin-boson model using the GKSL formulation.
 The closed system evolves under the Hamiltonian $\bm{H} = \hbar \omega \bm{S}_z/\sqrt{2}$, where $\hbar\omega$ is the spin energy splitting.
 The bath acts on the system through two Lindblad operators.
 The first, $\bm{L}_1 = \ket{1}\bra{2}$, induces excited to ground state transitions leading to dissipation with Lindblad rate $\Gamma$, while $\bm{L}_2 = \bm{S}_z$ fluctuates the spin energy splitting leading to dephasing with Lindblad rate $\gamma_\phi$.
Equation (\ref{eq:GKSL}) gives a flow field,
 \begin{equation}
   \label{eq:SBFlow}
   \dot{\bm{\sigma}} = [\omega y - (\gamma_\phi + \half\Gamma) x, -\omega x - (\gamma_\phi + \half\Gamma) y, \Gamma(1-z)],
 \end{equation}
 plotted in Fig. \ref{fig:q}.C.
 The unitary evolution gives a flow field, $\dot{\bm{\sigma}}_U = [\omega y, -\omega x, 0]$ (Fig. \ref{fig:q}.A), that precesses the spin about the z axis.
 Dissipative flow, $\dot{\bm{\sigma}}_{\textrm{Diss}} = [-\Gamma/2 x, -\Gamma/2 y, \Gamma(1-z)]$,  pushes the population into the ground state as the bath induces excited to ground state relaxation.
 Dephasing flow, $\dot{\bm{\sigma}}_{\textrm{Dep}} = [-\gamma_\phi x, -\gamma_\phi y, 0]$, does not drive any transitions and so acts perpendicular to $z$.
 However, the fluctuating energy gap destroys phase information driving pure states on the surface of the Bloch sphere to mixed states on the $z$ axis.
 When combined, this precesses the spin about the $z$ axis while spiraling towards the ground state.

Substituting Eq. (\ref{eq:SBFlow}) into Eq. (\ref{eq:localP}), closed systems with $\gamma_\phi = 0 =\Gamma$ have a vanishing compressibility, validating the quantum Liouville theorem.
For open systems, this yields the expected uniform compressibility $\kappa(\bm{\sigma}) = 2(\gamma_\phi + \Gamma)$, indicating that GKSL dynamics extend linear dissipation to quantum systems.
 In particular, the purely quantum dephasing process describing the loss of quantum phase information is formulated equivalently to classical friction.

 The dynamics of this Markovian Spin-Boson model are analytically known for all initial conditions $\bm{\sigma}_0 := [x_0, y_0, z_0]$.
This gives
 \begin{equation}
   \bm{\sigma}(t|\bm{\sigma}_0) = 
     \begin{bmatrix}
     \exp(-(\gamma_\phi+\Gamma/2)t)(x_0\cos(\omega t) - y_0\sin(\omega t))\\
     \exp(-(\gamma_\phi+\Gamma/2)t)(y_0\cos(\omega t) + x_0\sin(\omega t))\\
     \exp(-\Gamma t)(1 + z_0) -1
   \end{bmatrix}.
 \end{equation}
 The closed system trajectory (Fig. \ref{fig:q}.B) simply precesses about the $z$ axis while the open system (Fig. \ref{fig:q}.D) spirals towards the ground state. 

Combining Eqs. (\ref{eq:gfsol}) and (\ref{eq:SBFlow}), an initial distribution $P(\bm{\sigma}; 0)$ can be directly propagated.
For simplicity, consider a truncated Gaussian distribution, $P(\bm{\sigma}; 0)\propto \mathcal{N}(\bm{\sigma}; \bar{\bm{\sigma}}_0, \hat{\Delta}_0)\Theta(1-|\bm{\sigma}|)$ with initial mean $\bar{\bm{\sigma}}_0$ and covariance matrix $\hat{\Delta}_0$ and $\Theta$ is the Heaviside function.
The resulting time dependent distribution,
\begin{equation}
  \label{eq:PSB}
  P(\bm{\sigma}; t) \propto \mathcal{N}(\bm{\sigma}; \bar{\bm{\sigma}}(t| \bar{\bm{\sigma}}_0),\hat{\Delta}_t)\Theta(1- |\mathcal{S}(\bm{\nu}(t))\bm{\sigma}|)
\end{equation}
remains a truncated Gaussian at all times.
The time-dependent mean $\bar{\bm{\sigma}}(t|\bar{\bm{\sigma}}_0)$ is obtained by propagating the initial mean using Eq. (\ref{eq:SBFlow}).
The time dependent covariance matrix $\Delta_t = \mathcal{S}(\bm{\nu}(t))\mathcal{R}_z(\omega t) \Delta_0\mathcal{R}_z(\omega t)^T\mathcal{S}(\bm{\nu}(t))$  where $\mathcal{R}_z(\theta)$ is the z rotation by angle $\theta$ and $\mathcal{S}(\bm{\nu}(t))$ is the scaling matrix that scales the Cartesian axes by $\nu_x = \exp(-(\gamma_\phi + \Gamma/2)t)=\nu_y$, and $\nu_z = \exp(-\Gamma t)$.
This yields a Gaussian that rotates around the z axis while collapsing towards the ground state as shown in Fig. \ref{fig:q}.D.
The closed system distribution, shown in Fig. \ref{fig:q}.B, simply precesses about the $z$ axis.
We provide the solution for a general initial distribution in the SI.

In conclusion, the Liouville space of density operators provides a natural state space for the study and interpretation of quantum ensembles.
Individual quantum states are represented by discrete points in this state space and their dynamics obey flow properties identical to those of classical systems in phase space.
By exploiting the familiar classical structure of this quantum state-space it is possible to directly apply tools from classical statistical mechanics to quantum systems and to exercise classical intuition when interpreting their statistical properties.  
The formalism we have presented here thus casts the challenging problem of quantum mechanical mixed states in a form that is mathematically similar to classical ensembles, potentially enabling a unified treatment of quantum and classical statistical mechanics.

\onecolumngrid
\appendix 
\section{Glossary of Quantum-Classical Analogs}
\label{sec:QCA}

The Liouville state space formalism presented in the main text is constructed to mirror the structure of classical statistical mechanics on Hamiltonian phase space.
As such, nearly all objects in the classical theory have a quantum analog in the Liouville space theory.
Below, we provide a glossary that summarizes the key analogs between the two theories.

\begin{center}
  \begin{tabular}{c|c}
    \hline
    \textbf{Classical}  & \textbf{Quantum} \\
    \hline \hline
    Position-Momentum Phase Space: $\mathbb{R}^{2N} $ & Liouville State Space: $\Liou(\Hilb)$ \\ \hline
    State: $\bm{x} = [\bm{q}, \bm{p}]$ & State: $\bm{\sigma} = \hat{\sigma}$ \\ \hline
    Initial State: $\bm{x}_0$ & Initial State: $\bm{\sigma}_0$ \\ \hline
    Evolved State: $\bm{x}(t| \bm{x}_0)$& Evolved State:  $\bm{\sigma}(t|\bm{\sigma}_0)$\\ \hline
    Trajectory: $\lbrace \bm{x}(t| \bm{x}_0)| t \geq 0\rbrace$ &  Trajectory: $\lbrace \bm{\sigma}(t| \bm{\sigma}_0)| t \geq 0\rbrace$ \\ \hline 
    Dynamical Flow Field: $\dot{\bm{x}}$ & Dynamical Flow Field: $\dot{\bm{\sigma}}$\\ \hline
    Hamilton's Equations: $\dot{\bm{x}} = [\frac{\partial H}{\partial \bm{p}}, - \frac{\partial H}{\partial {\bm{q}}}]$ & Liouville-von Neumann Equation: $\dot{\bm{\sigma}} = -\frac{i}{\hbar}[\bm{H}, \bm{\sigma}]$ \\ \hline
    Linear Dissipator: $\hat{\Gamma}_\alpha$  & Lindblad Dissipator: $\mathcal{D}[\bm{L}_\alpha$] \\\hline
    Flow Compressibility: $\kappa(\bm{x}) = -\dive(\dot{\bm{x}})$ & Flow Compressibility $ \kappa(\bm{\sigma}) = -\dive(\dot{\bm{\sigma}})$ \\ \hline
    Continuity Equation: $\frac{\partial P}{\partial t} = -\bm{\nabla} P \cdot \dot{\bm{x}} + \kappa P$ &  Continuity Equation: $\frac{\partial P}{\partial t} = -\bm{\nabla} P \cdot \dot{\bm{\sigma}} + \kappa P$ \\ \hline
    Liouville's Equation (Closed): & Quantum Distribution Dynamics (Closed): \\
    $\frac{\partial P}{\partial t} = - \bm{\nabla} P \cdot [\frac{\partial H}{\partial \bm{p}}, - \frac{\partial H}{\partial {\bm{q}}}]$ &  $ \frac{\partial P}{\partial t} = \frac{i}{\hbar}\bm{\nabla} P \cdot[\bm{H}, \bm{\sigma}]$\\\hline
    Liouville Theorem (Closed): $\frac{d P}{d t}(\bm{x}(t); t_0)= 0$ & Liouville Theorem (Closed):  $\frac{d P}{d t}(\bm{\sigma}(t); t_0)= 0$ \\ \hline
    Liouville Theorem (Open):  & Liouville Theorem (Open):\\
    $\frac{d P}{d t}(\bm{x}(t); t_0)= \sum \Tr\hat{\Gamma}_\alpha \geq 0$ &  $\frac{d P}{d t}(\bm{x}(t); t_0)= \sum \gamma_\alpha \geq 0$ \\ \hline 
  \end{tabular}
\end{center}

\section{Non-Markovian Quantum Dynamical Flow}
\label{sec:NZE}

To derive an equation of motion for open quantum systems we follow the Nakajima-Zwanzig approach \cite{breuer_theory_2007}.
In this approach, the Liouville-von Neumann dynamics of a composite system and bath are projected (via the use of projection super-operators) onto two different Hilbert spaces called the relevant and irrelevant Hilbert spaces.
We define the relevant projection super-operator, $\Pproj$, as,
\begin{equation}
   \label{eq:Pproj}
   \Pproj\bm{\sigma}=\TrB\left\{\bm\sigma\right\}\otimes\bm{\rho_\mathrm{B}},
 \end{equation}
where $\TrB$ indicates a trace over the bath and $\bm{\rho_\mathrm{B}}\in \Liou ( \Hilb_\mathrm{B})$ is a stationary bath reference state that is normalized so that $\Tr_B\left(\bm{\rho_\mathrm{B}}\right)=1$.
Defined in this way, $\Pproj \bm{\sigma}$ yields a projected density operator $\bm{{\sigma}_\mathrm{rel}} \in \Liou\left(\Hilb_S\otimes\Hilb_B\right)$, that is related to the more familiar reduced system density operator via $\bm{\sigma_\mathrm{S}}=\TrB\left(\bm{\sigma_\mathrm{rel}}\right)$.
The irrelevant projection super-operator is given by,
\begin{equation}
    \label{eq:Qproj}
    \Qproj=\mathbb{I}-\Pproj,
\end{equation}
where $\mathbb{I}$ is the identity super-operator.

With these projection super-operators, the dynamics of the irrelevant space can be formally solved and expressed in terms of its effect on the dynamics within the relevant Hilbert space. 
Specifically, consider a general system-bath Hamiltonian,
\begin{equation}
\bm{H}=\bm{H_\mathrm{S}} + \bm{H_\mathrm{B}} + \bm{V},
\label{eq:sysbath}
\end{equation}
where $\bm{H_\mathrm{S}}$ is the system Hamiltonian, $\bm{H_\mathrm{B}}$ is the bath Hamiltonian, and $\bm{V}$ describes the interaction between the system and bath.
The dynamics of such a system can be expressed in the interaction picture using the Nakajima-Zwanzig equation,
\begin{subequations}
  \label{eqs:NJE}
  \begin{equation}
    \label{eq:NJE}
    \frac{\partial \bm{\sigma_{rel}}}{\partial t}=\Pproj\Lprop(t)\bm{\sigma_{rel}}(t)+\Pproj\Lprop(t)\Gprop(t,t_0)\Qproj\bm{\sigma_{rel}}(t_0)+\int^t_{t_0}ds\K(t,s)\bm{\sigma_{rel}}(s),
  \end{equation}
where,
  \begin{equation}
    \label{eq:Kern}
    \K(t,s):=\Pproj\Lprop(t)\Gprop(t,s)\Qproj\Lprop(s)\Pproj,
  \end{equation}
 and,
  \begin{equation}
    \label{eq:Gprop}
    \Gprop(t,s):=\mathcal{T}_\leftarrow \exp\left[\int^t_s ds' \Qproj \Lprop(s')\right],
  \end{equation}
\end{subequations}
where $\Lprop(t)\bm{\sigma}:=\bm{\left[{V}(t),{\sigma}(t)\right]}$ is the Liouville super-operator, $\mathcal{T}_\leftarrow$ is the time-ordering superoperator, and $t_0$ corresponds to the time at which the system is initialized.

The Nakajima-Zwanzig equation can be simplified with the appropriate choice of initial conditions.
For example, the first term in Eq.~\ref{eq:NJE} vanishes in the case where the bath reference state is selected so that $\TrB\left\{\bm{V(t)\rho_B}\right\}=\bm{0}$ \cite{breuer_theory_2007}.
Likewise, the second term (describing contributions arising due to entangled initial conditions) vanishes when $\bm{\sigma}(t_0)=\bm{\sigma_S}(t_0)\otimes\bm{\rho_B}$.
For simplicity, we will restrict out attention to systems whose initial conditions cause the first two terms in Eq.~(\ref{eq:NJE}) to vanish in this way.
This leaves only a homogeneous integro-differential equation, as described by the final term in Eq.~(\ref{eq:NJE}).

Dynamical flow under the Nakajima-Zwanzig equation can be determined from the non-Markovian flow field, $\bm{\dot{\sigma}_\mathrm{S}}=\TrB \lbrace \bm{\dot{\sigma}_\mathrm{rel}} \rbrace$.
To compute the divergence of this flow field, we take the component-wise functional derivative of $\bm{\dot{\sigma}_\mathrm{S}}$ with respect to $\bm{\sigma_\mathrm{S}}$ to get,
\begin{equation}
  \label{eq:diagonal}
  \frac{\delta[\dot{\sigma}_{S;i,j}(t)]}{\delta[\sigma_{S;i,j}(t')]}=\sum_\alpha\sum_{\delta,\gamma}\mathcal{K}_{i\alpha,j\alpha;i\delta,j\gamma}(t,t')\rho_{B;\delta,\gamma},
\end{equation}
where the bath is referred to with Greek indices and the system is referred to with Latin indices. 

The divergence can be computed from this expression to yield, 
\begin{equation}
  \label{eq:NJcomp}
  \bm{\nabla}\cdot\dot{\bm{\sigma}}_S(t)=\sum_{i,j}\sum_{\alpha, \delta, \gamma}\mathcal{K}_{i\alpha,j\alpha;i\delta,j\gamma}(t,t)\rho_{B;\delta,\gamma},
\end{equation}
which represents the compressibility of Nakajima-Zwanzig flow on the state space of reduced density operators. 
This expression corresponds to general non-Markovian dynamics, including the Markovian limit where the memory kernel  $ \K (t,t')\propto \delta(t-t')$.
Notably, a similar functional derivative approach is also used derive the Euler-La Grange Equation.
Finally, combining Eq. (\ref{eq:NJcomp}) and the fluid mechanical equation of motion in the main text, leads to an equation of motion for open quantum systems of the form,
\begin{equation}
\label{eq:openEOM}
\frac{\partial P}{\partial t} (\bm{\sigma}; t)= \frac{i}{\hbar} \bm{\nabla}P(\bm{\sigma}; t)\cdot\TrB\left\{\frac{\partial\bm{\sigma}_{rel}}{\partial t}\right\} + P(\bm{\sigma}; t)\sum_{i,j}\sum_{\alpha, \delta, \gamma}\mathcal{K}_{i\alpha,j\alpha;i\delta,j\gamma}(t,t)\rho_{B;\delta,\gamma},
\end{equation}
which is similar to that of the generalized Langevin equation \cite{zwanzig_nonequilibrium_2001}.
This formalism can thus be used to develop a quantum analog to the generalized non-Markovian Fokker Planck equation.

\section{Derivation of the Quantum Continuity Equation}
\label{sec:QCE}

The quantum continuity equation (Eq. (4) in the main text) is one of the key results of this paper.
This expression relates the dynamics of single subsystems, e.g. governed by the Liouville von-Neumann, GKSL or Nakajima-Zwanzig equation, to the dynamics of the distribution.
It takes a form analogous to the classical continuity equation and can be derived in an indentical manner.
To derive the continuity equation, we consider the probability $P_\Omega(t)$ of finding a system in a region $\Omega$ at time $t$.
This quantity can change in one of 3 ways.
(1) A subsystem state can enter or leave $\Omega$. That is there exists a time $t'$ where some subsystem trajectory is outside of $\Omega$ immediately before $t'$ and inside it immediately after.
(2) A subsystem is created in $\Omega$. That is at a time $t'$ a new trajectory is created inside $\Omega$ that was not defined for $0\leq t < t'$.
(3) A subsystem is destroyed in $\omega$. That is at a time $t'$ a trajectory that was inside $\Omega$ is no longer defined for $t > t'$.

In all forms of dynamics considered in this paper, the evolution of a subsystem state is defined for all time $t \geq 0$, since a system can be propagated by specifying an initial state $\bm{\sigma}_0$.
This property of the dynamics prevents the creation or destruction of trajectories required for processes (2) and (3) above to occur since they require a trajectory to be undefined for some times $t \geq 0$.
Trajectories generated by these dynamics are said to be conserved and all change in $P_\Omega$ must therefore occur due to a subsystem entering or leaving $\Omega$ through process (1).
Moreover, since all trajectories produced by these dynamics are continuous, a trajectory cannot enter or leave $\Omega$ without passing through it's boundary.

Consequently, the change in $P_\Omega$ must be related to the probability flux, $\bm{j}(\bm{\sigma}; t) := P(\bm{\sigma}; t)\dot{\bm{\sigma}}(t)$, at it's boundary, giving the integral form of the continuity equation
\begin{equation}
  \label{eq:intCE}
  \frac{d P_\Omega}{d t} := \frac{d} {dt} \int_{\bm{\sigma}\in\Omega} P(\bm{\sigma}; t) d\bm{\sigma} = -\int_{\bm{\sigma}\in \partial\Omega}\bm{j}(\bm{\sigma}; t) \cdot d\bm{A}(\bm{\sigma}),
\end{equation}
where the second integral is taken over the boundary, $\partial \Omega$, of $\Omega$ and $d\bm{A}(\bm{\sigma})$ denotes the infinitessimal surface normal at $\bm{\sigma}$ in the surface-flux integral.

Equation (\ref{eq:intCE}) can be simplified to give the differential form reported in the main text by applying Gauss's Divergence theorem to the surface-flux integral.
This gives
\begin{equation}
  \label{eq:Gauss}
   \int_{\bm{\sigma}\in \partial\Omega}\bm{j}(\bm{\sigma}; t) \cdot d\bm{A}(\bm{\sigma}) = \int_{\bm{\sigma}\in \Omega}\dive(\bm{j}(\bm{\sigma}; t)) d\bm{\sigma},
\end{equation}
which applies to any region $\Omega$.
By taking $\Omega$ to be an infinitessimal volume element surrounding a point $\bm{\sigma}$ (e.g. as the limit as $r \to 0$ of open balls with radius $r$ centered at $\bm{\sigma}$), we then obtain the differential form
\begin{equation}
  \label{eq:QCE}
  \frac{\partial P}{\partial t} (\bm{\sigma}; t) = -\dive(\bm{j}(\bm{\sigma}; t)) = -\bm{\nabla} P(\bm{\sigma}; t) \cdot \dot{\bm{\sigma}} + \kappa(\bm{\sigma}; t) P(\bm{\sigma}; t),
\end{equation}
where we have applied the divergence product rule to $\bm{j}(\bm{\sigma}; t)$ and the compressibility  $\kappa(\bm{\sigma}; t) := - \dive(\dot{\bm{\sigma}}(t)$ is taken to be time dependent to account for the time dependent flow fields that can arise in time convolutionless expressions of Nakajima-Zwanzig dynamics.

\section{Derivation of Spin-Boson Distribution Dynamics}
\label{sec:SBD}

We now consider the distribution dynamics of a spin system coupled to a Bosonic bath.
It is convenient to express the dynamics of the system in the Pauli Basis $\lbrace \bm{S}_i \rbrace_{i\in\lbrace0, x, y, z\rbrace}$ where $\bm{S}_0 := \bm{\mathbb{I}}/\sqrt{2}$ is the normalized identity matrix and 
$\bm{S}_x := (\ket{1}{2} + \ket{2}{1})/\sqrt{2}$, $\bm{S}_y := -i(\ket{1}\bra{2} -\ket{2}\bra{1})/\sqrt{2}$, and $\bm{S}_z := (\ket{2}\bra{2} - \ket{1}\bra{1})/\sqrt{2}$ are the normalized Pauli matrices.
The $1/\sqrt{2}$ normalization constant comes about due to normalization with respect to the trace inner product.
In this notation, the system Hamiltonian is given by 
\begin{equation}
  \label{eq:SBHam}
  \bm{H} = \frac{\hbar \omega}{\sqrt{2}} \bm{S}_z,
\end{equation}
and the density matrix can be conveninetly written in Bloch vector notation as $\bm{\sigma}:= (x\bm{S}_x +y\bm{S}_y + z\bm{S}_z + \bm{\mathbb{I}})/\sqrt{2} \to [x, y, z].$
The bath interactions are treated phenomenologically through the following Lindblad Dissipators:
\begin{subequations}
  \begin{equation}
    \label{eq:D_diss}
    \mathcal{D}[\bm{L_1}]\bm{\sigma} =
    \begin{bmatrix}
      -\half x\\
      -\half y\\
      -(1 + z)
    \end{bmatrix}
  \end{equation}
  \begin{equation}
    \label{eq:D_dep}
    \mathcal{D}[\bm{L_2}] \bm{\sigma} =
    \begin{bmatrix}
      -x\\
      -y\\
      0
    \end{bmatrix}
  \end{equation}
\end{subequations}
where $\bm{L_1} := \ket{1}\bra{2}$ is the dissipation operator that induces transitions from the excited state to the ground state and $\bm{L_2}:= \bm{\sigma_z}$ is the pure dephasing operator that fluctuates the spin energy splitting.
The Lindblad rates for these processes are given by the dissipation rate $\Gamma$ and the dephasing rate $\gamma_\phi$ respectively.

Using the GKSL equation (Eq. (3) of the main text) the dynamical flow field for this model system can then be written as
\begin{equation}
  \label{eq:SBFlow}
  \dot{\bm{\sigma}} =
  \begin{bmatrix}
    -\omega y -\left(\gamma_\phi + \half\Gamma \right)x\\
    \omega x - \left(\gamma_\phi + \half\Gamma \right)y \\
    -\Gamma(1+ z)
  \end{bmatrix}.
\end{equation}
The subsystem dynamics generated by Eq. (\ref{eq:SBFlow}) are analytically solvable.
The dynamics are given by
\begin{equation}
  \label{eq:SBtraj}
  \bm{\sigma}(t|\bm{\sigma}_0) = 
    \begin{bmatrix}
    \exp(-(\gamma_\phi+\Gamma/2)t)(x_0\cos(\omega t) - y_0\sin(\omega t))\\
    \exp(-(\gamma_\phi+\Gamma/2)t)(y_0\cos(\omega t) + x_0\sin(\omega t))\\
    \exp(-\Gamma t)(1 + z_0) -1
  \end{bmatrix},
\end{equation}
where the initial condition is written in Bloch vector notation as $\bm{\sigma}_0 := [x_0, y_0, z_0]$.

These subsystem results can be generalized to obtain the dynamics of an arbitrary initial distribution $P(\bm{\sigma}; 0)$ using  convective Green's Function equation given by main text Eq. (6) as
\begin{equation}
  \label{eq:GF}
  P(\bm{\sigma}; t) = \int d\bm{\sigma'} P(\bm{\sigma'}; 0) \delta(\bm{\sigma}(t|\bm{\sigma'})- \bm{\sigma}),
\end{equation}
where the convective Green's function $ \delta(\bm{\sigma}(t|\bm{\sigma'})- \bm{\sigma})$ selects initial conditions , $\bm{\sigma'} =[x', y', z']$ that evolve to $\bm{\sigma}$ at time $t$.
Using Eq. (\ref{eq:SBtraj}), the convective Green's Function can be simplified to give
\begin{equation}
  \label{eq:GF_explicit}
  \begin{aligned}
    \delta(\bm{\sigma}(t|\bm{\sigma'})- \bm{\sigma}) = &\delta\left(x'-e^{(\gamma_\phi + \half\Gamma)t}(x\cos(\omega t) + y\sin(\omega t))\right) \times \\
    & \delta\left(y' - e^{(\gamma_\phi +\half\Gamma)t}(y\cos(\omega t) - x\sin(\omega t) )\right) \times\\
    & \delta\left(z' - e^{\Gamma t}(1 + z) - 1 \right),
    \end{aligned}
\end{equation}
which expresses the Green's function as explicit functions of the integrating variable $\bm{\sigma'}$.

Combining Eqs. (\ref{eq:GF}) and (\ref{eq:GF_explicit}), the time-dependent distribution can be found by transforming the coordinates of the initial distribution $P(\bm{\sigma}; 0)$ giving
\begin{subequations}
  \label{eqs:tdist}
  \begin{equation}
    \label{eq:tdist}
    P([x, y, z]; t) = e^{2(\gamma +\Gamma) t}P(\tilde{\bm{\sigma}}_t; 0)
  \end{equation}
  \begin{equation}
    \tilde{\bm{\sigma}}_t = \mathcal{S}([e^{(\gamma_\phi +\half\Gamma)t},e^{(\gamma_\phi +\half\Gamma)t}, e^{\Gamma t}]) \mathcal{R}_z(-\omega t) \bm{\sigma}
  \end{equation}
\end{subequations}
where $\mathcal{S}(\bm{\nu})$ is the scaling matrix and $\mathcal{R}_z(\theta)$ is the matrix for a rotation by angle $\theta$ about the $z$ axis of the Bloch sphere.
The normalization factor in Eq. (\ref{eq:tdist}) accounts for the change in normalization due to the scaling transformation and arises from the determinant of the scaling transformation.
These transformations are explicitly written in matrix form as
\begin{subequations}
  \label{eqs:transforamtions}
  \begin{equation}
    \label{eq:scaling}
    \mathcal{S}(\bm\nu) :=
    \left(
    \begin{matrix}
      \nu_x & 0 & 0 \\
      0 & \nu_y & 0 \\
      0 & 0 & \nu_z
    \end{matrix}
    \right)
  \end{equation}
  \begin{equation}
    \label{eq:rotation}
    \mathcal{R}_z(\theta) :=
    \left(
    \begin{matrix}
      \cos\theta & -\sin\theta & 0 \\
      \sin\theta & \cos\theta & 0 \\
      0 & 0 & 1
    \end{matrix}
    \right).
  \end{equation}
\end{subequations}

In the main text, we consider a Gaussian initial distribution with mean $\bar{\bm{\sigma}}_0$ and covariance matrix $\hat{\Delta}_0$ given by
\begin{equation}
  \label{eq:gaussian}
  \mathcal{N}(\bm{\sigma}; \bar{\bm{\sigma}}, \hat{\Delta}) := \frac{1}{(2\pi)^{\frac{3}{2}}\left|\hat{\Delta}\right|^\half} \exp\left(-\half(\bm{\sigma} - \bar{\bm{\sigma}})\cdot\hat{\Delta}^{-1}(\bm{\sigma}-\bar{\bm{\sigma}})\right)
\end{equation},
where $\left|A\right|$ denotes the determinant of $A$ and $A^{-1}$ it's inverse.
Applying Eq. (\ref{eqs:tdist}) to a Gaussian initial distribution yields a Gaussian at all times with time varying mean and covariance
\begin{equation}
  \label{eq:tgauss}
  P(\bm{\sigma}; t) = \mathcal{N}(\bm{\sigma}; \bar{\bm{\sigma}}(t|\bar{\bm{\sigma}}_0), \hat{\Delta}_t),
\end{equation}
where $\bar{\bm{\sigma}}(t|\bar{\bm{\sigma}}_0)$ is obtained by propagating the initial mean using Eq. (\ref{eq:SBtraj}) and the time dependent covariance is given by the transformation $\hat{\Delta}_t := \mathcal{S}(\bm{\nu}(t)) \mathcal{R}_z(\omega t) \hat{\Delta}_0 \mathcal{R}_z(\omega t)^T \mathcal{S}(\bm{\nu}(t))$ where the scaling vector is given by $\bm{\nu(t)} := [e^{-(\gamma_\phi +\half\Gamma)t},e^{-(\gamma_\phi +\half\Gamma)t}, e^{-\Gamma t}]$.
Notably, the scaling factor in Eq. (\ref{eq:tdist}) accounts for the renormalization of the Gaussian due to the time varying covariance matrix.

\bibliography{state_space}

\end{document}